\pdfoutput=1

\documentclass[reprint, amssymb, amsmath, aip, apl]{revtex4-1}

\usepackage{docs}
\usepackage{graphicx} 
\usepackage{dcolumn}
\usepackage{bm}
\usepackage{epstopdf}

\begin{document}

\title{Remote sensing and control of phase qubits}


\author{Dale Li}
\email{dale.li@boulder.nist.gov}
\author{Fabio C.S. da Silva}
\author{Danielle A. Braje}
\author{Raymond W. Simmonds}
\author{David P. Pappas}
\affiliation{National Institute of Standards and Technology, Boulder, Colorado 80305, USA}


\date{\today}

\begin{abstract} 
We demonstrate a remote sensing design of phase qubits by separating the control and readout circuits from the qubit loop.  This design improves measurement reliability because the control readout chip can be fabricated using more robust materials and can be reused to test different qubit chips.  Typical qubit measurements such as Rabi oscillations, spectroscopy, and excited-state energy relaxation are presented.
\end{abstract}


\maketitle


Superconducting phase qubits are one of the most promising technologies for a scalable quantum computer.\cite{MartinisQIP}  Introduction and improvement of specialized materials and structures has significantly reduced losses and improved coherence times.\cite{Oh} However, evaluation of these materials creates challenges in the design and fabrication of qubit circuits primarily because of variations in material composition and crystalline order.\cite{Kline} 
The ability to explore different materials would be greatly simplified if the control and readout circuit to measure the qubit could be fabricated separately from the qubit devices under investigation.  The readout circuit could then be made of well-established materials and designs, and would operate reliably independent of materials being used for the qubits.  In this letter, we developed a self-aligning flip-chip technique to separate the qubit circuit from its readout.  The readout chip is inductively coupled to the phase qubit, and contains the SQUID readout and the superconducting coils for microwave and dc flux control.

Previous superconducting circuits have used flip-chips to perform noise and remote detection measurements \cite{BerggrenRemote, MayCoil}.  Flip-chip implementations of charge qubits operating as interferometers have also been reported.\cite{BornChargeQ} In addition, flip-chips have been used to separate dissipative single-flux quantum (SFQ) circuits from the temperature-sensitive qubit circuits.\cite{YorozuFlux}
Bennett {\it et al.\ }used a separate chip suspended above an rf-SQUID qubit chip to obtain fast bias pulses.\cite{BennettFlip}
Steffen {\it et al.} describe a SQUID-less readout scheme that reduces the number of junctions in the qubit to one (the qubit junction itself). This scheme allows for the multiplexing of many qubits. However, the overall performance of the system is affected by the coupling between the microwave feed line and the qubit circuit.\cite{DispReadout}
Michotte uses the flip-chip technique to separate the microstrip line from the SQUID sensor in a 
microstrip-SQUID amplifier.\cite{MicrostripFlip}

\begin{figure}[t]
\includegraphics[width=3.4in]{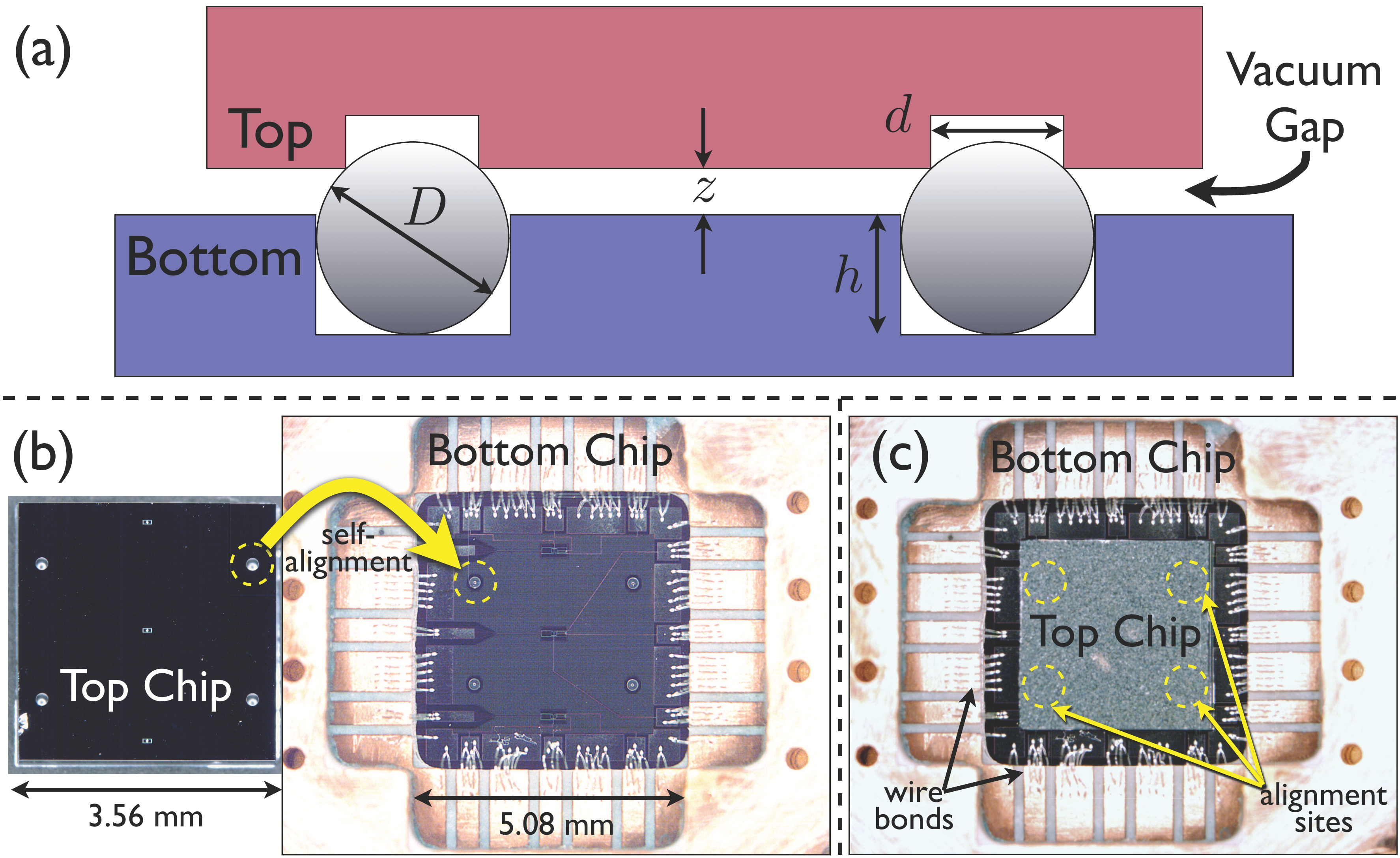}
\caption{\label{fig1}(color online).
(a) The top chip is self-aligned a distance $z$ above the bottom chip by use of sapphire spheres of diameter $D$.  The top chip pocket diameter $d$ is determined by Eq.\ (1) with a fixed bottom chip pocket depth of $h$. (b) The top chip and bottom chip are separated, showing the alignment sites and the scale of each chip.  Note that the top chip is smaller than the bottom chip to allow space for wire bonding.  (c) The assembled flip-chip.
}
\end{figure}

Our flip-chip design contains the phase qubit loop on the top chip, which self-aligns, by use of four $200 \pm 2.5$ $\mu$m diameter sapphire spheres, to the bottom chip containing the control/readout circuitry.  Sapphire spheres have a small thermal contraction coefficient, which helps to maintain proper alignment when the sample is cooled to dilution-refrigerator temperatures.  The spheres sit in pockets etched into the silicon substrates by a deep reactive ion etcher.  

Figure 1(a) shows a cross-sectional drawing of the deeply etched cylindrical pockets in the top and bottom chips and the self-aligning sapphire spheres.  The diameter of the top chip pocket is given by $d = 2\sqrt{(D-h-z)(h+z)}$, where $D$ is the diameter of the sapphire sphere, $h$ is the depth of the pocket etched into the bottom chip (with etched diameter equal to $D$), and $z$ is the desired vacuum gap size.  
Deep pockets in the bottom chip held the sapphire spheres in place for reuse, while the shallower pockets in the top chip were etched deep enough that the sapphire spheres only touch the top chip at the edges of the pockets.
Different pocket diameters for different top chips were fabricated, giving vacuum gap sizes from 10 $\mu$m to 50 $\mu$m.  

Photographs of the fabricated top and bottom chips are shown separately, with the bottom chip wire-bonded to a test board in Fig.\ 1(b), and in the flip-chip configuration in Fig.\ 1(c).  The four positions for the sapphire spheres facilitate a stable self-alignment, minimize wobble, and place the spheres far away from the circuit elements.    
The entire flip-chip assembly is held together under slight compression by a beryllium-copper leaf spring placed inside a brass lid, which encloses the two chips and fastens to the circuit board.

\begin{figure}[t]
\includegraphics[width=3.2in]{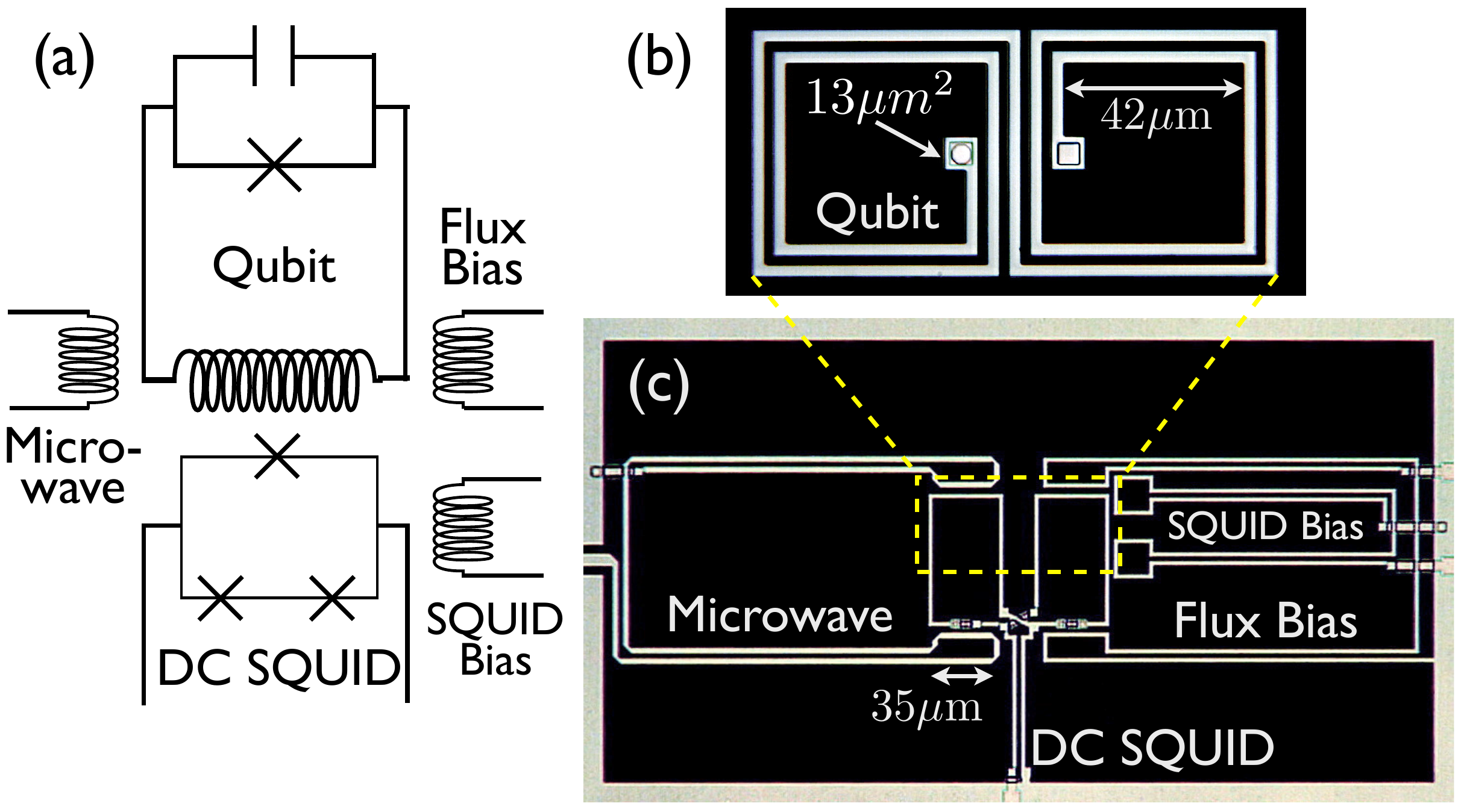}
\caption{\label{fig2}(color online).
(a) Flip-chip circuit drawing shows the simple qubit circuit and the three inductively coupled control coils for microwave excitation, DC flux bias, and DC SQUID bias, as well as the DC three-junction SQUID for qubit readout.  (b) A photograph of qubit loop near the final steps of fabrication as patterned on the top chip.  A final wiring layer connects the junction and the via (not shown).  (c) Photograph of measurement and excitation circuitry as fabricated on the bottom chip.  The dashed large rectangle indicates where the qubit will align.
}
\end{figure}

Figure 2(a) shows the circuit model for the entire phase qubit including control and readout (C/R).  The C/R circuit consists of a three-junction dc SQUID (readout), a dc flux bias loop that applies magnetic flux to the qubit (control), a secondary dc flux bias loop to tune the magnetic flux in the SQUID (control), and a microwave flux loop that excites the qubit with microwave frequencies (control).  Each inductive loop utilizes a gradiometric design to minimize  both unwanted cross-coupling between coils and the effects of shifts in background homogeneous magnetic fields by symmetric placement. Fig.\ 2(b) shows a photograph of the qubit loop as patterned on the top chip.  To test this flip-chip approach, standard Al/amorphous-Al$_2$O$_3$/Al Josephson junctions $13$ $\mu$m$^2$ in area were designed and fabricated for qubit frequencies around 7 GHz.  The qubit loop was closed by an Al cross-over wire connecting the junction and the via (not shown).  Fig.\ 2(c) shows a photograph of the C/R circuitry above which the qubit loop is placed (dashed rectangle) when aligned.

For a vacuum gap size $z=20$ $\mu$m, the mutual inductance coupling terms were calculated between pairs of coils (qubit-SQUID: 71 pH, qubit-flux bias: 5.5 pH, qubit-SQUID bias: $<$1 pH, qubit-microwave line: 5.5 pH, SQUID-SQUID bias: 2 pH, SQUID-flux bias: $<$1 pH).  The qubit loop was designed with a self inductance of 880 pH, while the SQUID was designed with a self-inductance of 341 pH.
These large inductances ensured a strong measurable coupling between the qubit chip and the C/R chip, although smaller inductances could also provide adequate coupling, depending on the gap size.

\begin{figure}[t]
\includegraphics[width=3.1in]{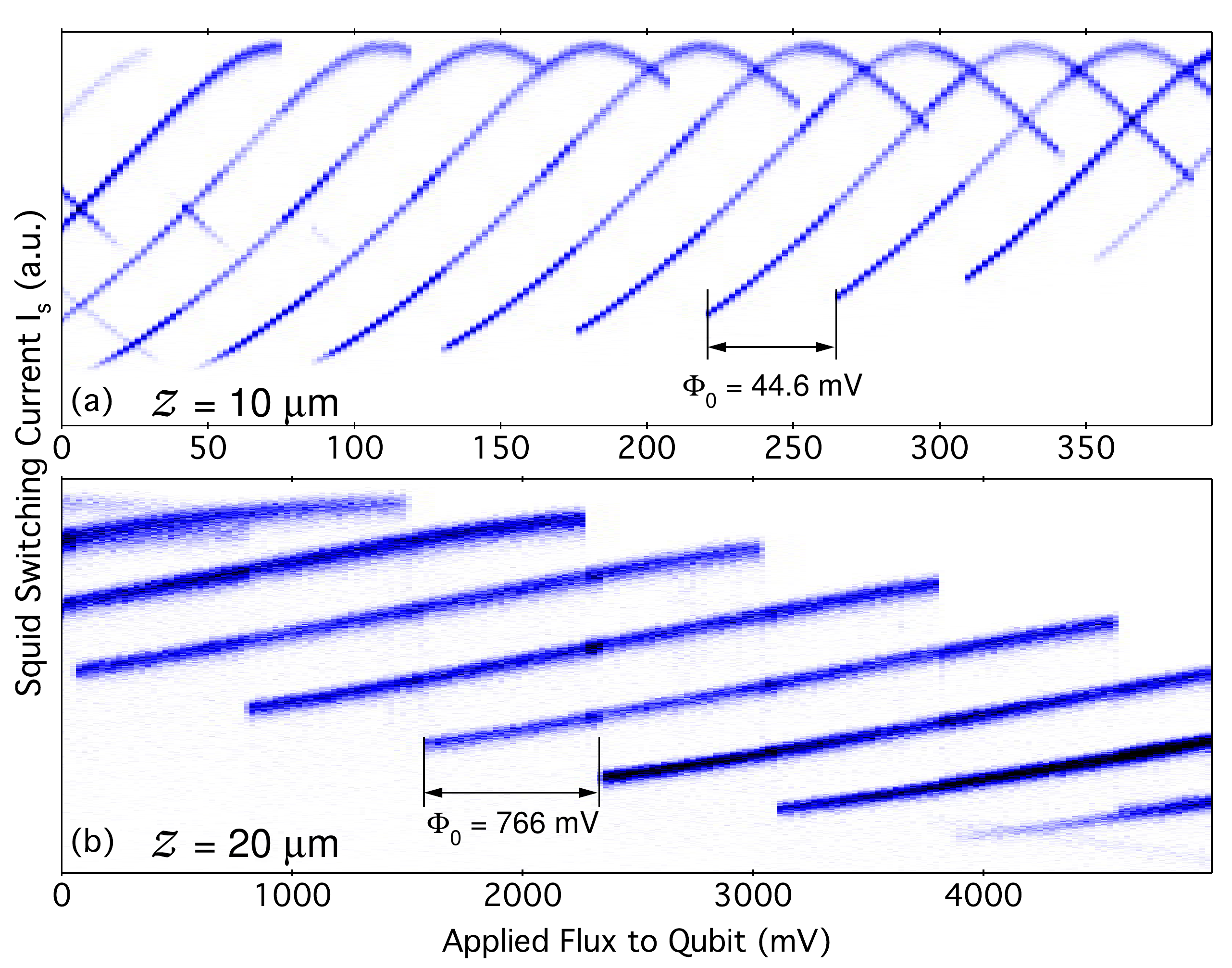}
\caption{\label{fig3}(color online).
Qubit steps for two different qubit chips (same readout chip) showing different coupling.  (a) $z$=$10$ $\mu$m gap size.  The steps are curved due to the large overlap coupling to the dc SQUID.  A flux quantum in the qubit is observed with the applied voltage $\Phi_0$=$44.6$ mV. (b) $z$=$20$ $\mu$m gap size has weaker coupling and samples just the linear regime of the SQUID. A larger applied voltage is needed to excite a flux quantum with $\Phi_0$=$766$ mV.
}
\end{figure}

We tested the remote sensing and control of the phase qubit with four typical measurements showing coherent control and reliable readout: qubit steps, spectroscopy, Rabi oscillations, and $T_1$.\cite{HistFitting}  Additionally, the response of the SQUID was measured as a function of the applied flux through the SQUID bias line in order to test the C/R circuit independently of the qubit.  The SQUID bias line also provided the ability to tune the SQUID to a sensitive, mostly linear regime.

First, we measured the qubit steps by applying a magnetic flux to the qubit loop
and measuring the corresponding value of the SQUID switching current $I_{s}$. Here,
the applied flux is measured in units of the voltage across a 10 k$\Omega$ resistor
connected in series with the qubit bias coil.
Figure \ref{fig3}(a) shows the behavior of $I_{s}$ versus the applied flux for a gap size
between the bottom and top chips of $10$ $\mu$m. The pronounced nonlinear
behavior of $I_{s}$ arises from a large field change as sensed by the SQUID at different qubit states, which maps to a larger, less linear regime in the SQUID response.   For this gap size of $10$ $\mu$m, the voltage difference necessary to induce a quantum of flux
($\Phi_{0}$) variation in the qubit is 44.6 mV. For an increased gap size
of $20$ $\mu$m, the flux bias voltage per flux quantum increased to 766 mV as shown in Fig.\ \ref{fig3}(b).  This change in flux bias per flux quantum corresponds to a reduction of the coupling by a factor of 17.
 Furthermore, the reduction of coupling decreased the amount of qubit flux sensed by the SQUID so that its response mapped to a more linear regime, as shown in Fig.\ \ref{fig3}(b).

\begin{figure}[t]
\includegraphics[width=3.1in]{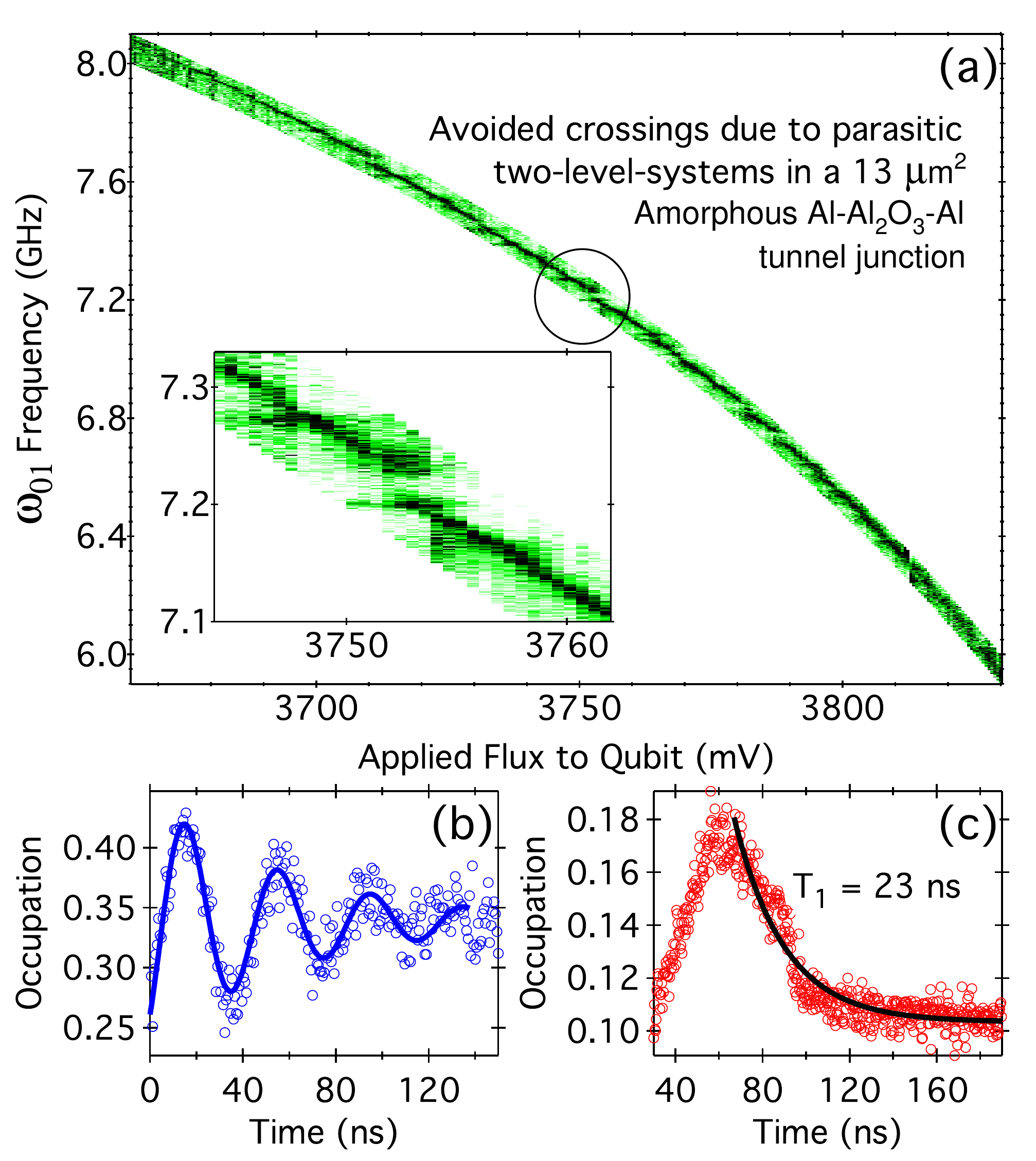}
\caption{\label{fig4}(color online).
Data collected from $z=20$ $\mu$m gap sized flip-chip. (a) Spectroscopy data showing the tunability of the qubit resonant frequency as a function of the applied flux from the bottom chip. The inset shows a zoom in of one of many splittings due to coupling with parasitic two level systems in this qubit. (b) Rabi Oscillations in the qubit from microwave excitation. (c) Relaxation time measurement.  
} 
\end{figure}

Second, we measured the qubit spectroscopy for a gap size of 20 $\mu$m.  The phase qubit exhibits a tunable absorption spectrum at its transition frequency ($\omega_{01}$) between the ground and first excited state.  In Fig.\ 4(a) the qubit spectroscopy shows a 2 GHz range of $\omega_{01}$ values centered around 7 GHz.  The visibility of only one transition line in the spectroscopy data indicates that the qubit chip was cooled to low enough temperatures to be operated as a qubit.   The discontinuities in the spectrum are assumed to be due to parasitic two-level systems in the large-area amorphous-Al$_2$O$_3$ tunnel barrier.\cite{Simmonds2004} A zoom-in of one such discontinuity is shown in the inset.  

Third, Fig.\ 4(b) shows Rabi oscillations in the same qubit.  This experiment is performed by holding a constant dc flux bias in a region of the spectroscopy with few discontinuities and applying a microwave pulse for a varied period. Rabi oscillations demonstrate the ability for state mixing between the ground and first excited state of the qubit.  The oscillation amplitude decays due to decoherence with a spin bath and should ideally saturate to a $50\%$ occupation probability.  In the data, the saturation occurs at about a $33\%$ occupation probability.  This discrepancy is due to the measurement process, which sweeps the coupling of the qubit through many avoided crossings with parasitic two-level systems that syphon energy from the qubit in Landau-Zener-like transitions.\cite{Cooper2004}

Fourth, Fig.\ 4(c) shows a longitudinal relaxation experiment in the same qubit.  In this experiment, a partially excited qubit state is prepared with a fixed microwave pulse length of 50 ns, and the qubit state is measured as a function of time as it decays to its ground state.
Our flip-chip test used similar design considerations, materials, and fabrication techniques as for integrated chips so we expected the experimental data to agree with previous results without the introduction of additional noise or loss.  Though the observed relaxation time $T_1=23$ ns is short, it matches reported results for a phase qubit with a $13$ $\mu$m$^2$ thermally oxidized amorphous Al$_2$O$_3$ tunnel barrier on a Silicon substrate.\cite{Cooper2004}

In conclusion, we demonstrated the remote sensing and control of a phase qubit by separating the qubit loop and the control/readout (C/R) circuit.  Typical characterization and performance measurements done in several qubit loops with the same C/R circuit demonstrated reliability and robustness of this design.  The technique has therefore proven to be an adequate candidate for studying the improvement of specialized materials and structures for superconducting qubits.
Other types of qubits, such as flux qubits could also potentially use the same flip-chip technique either by direct coupling across a smaller controlled gap, or by mediated coupling through a resonator circuit or rf-SQUID.\cite{Shane}

This research was funded in part by the Office of the Director of National Intelligence (ODNI) and by Intelligence Advanced Research Projects Activity (IARPA).  All statements of fact, opinion or conclusions contained herein are those of the authors and should not be construed as representing the official views or policies of IARPA, the ODNI, or the U.S. Government. Official contribution of the National Institute of Standards and Technology; not subject to copyright in the United States.



\begin{references}
\bibitem{MartinisQIP}J.M. Martinis, Quantum Inf. Process {\bf 8}, 81 (2009).
\bibitem{Oh}S. Oh, K. Cicak, J.S. Kline, M.A. Sillanpaa, K.D. Osborn, J.D. Whittaker, R.W. Simmonds, and D.P. Pappas, Phys. Rev. B {\bf 74}, 100502(R) (2006).
\bibitem{Kline}J.S. Kline, H. Wang, S. Oh, J.M. Martinis, and D.P. Pappas, Supercond. Sci. Technol. {\bf 22}, 015004 (2009).
\bibitem{BerggrenRemote}K.K. Berggren, D. Nakadab, M. J. O'Hara, T.P. Orlando, E.M. Macedo, R. Slattery, T. Weir,  {\it An Integrated Superconductive Device Technology for Qubit Control} (Rinton, Princeton, 2001), pp. 121-126.
\bibitem{MayCoil}T. May, E. Il'ichev, and H.-G. Meyer. Rev Sci. Instrum. {\bf 74}, 1282 (2003).
\bibitem{BornChargeQ}D. Born, V.I. Shnyrkov, W. Krech, Th. Wagner, E. Il'ichev, M. Grajcar, U. Hubner, H.-G.Meyer, Phys. Rev. B {\bf 70}, 180501(R) (2004).
\bibitem{YorozuFlux} S. Yorozu, T. Miyazaki, V. Semenov, Y. Nakamura, Y. Hashimoto, K. Hinode, T. Sate, Y. Kameda, and J.S. Tsai. J. Phys: Conf. Series {\bf 43}, 1417 (2006).
\bibitem{BennettFlip}D.A. Bennett, L. Longobardi, V. Patel, W. Chen, J.E. Lukens. Supercond. Sci. Technol. {\bf 20}, S445 (2007).
\bibitem{DispReadout}M. Steffen, S. Kumar, D. DiVincenzo, G. Keefe, M. Ketchen, M.B. Rothwell, and J. Rozen. Appl. Phys. Lett. {\bf 96}, 102506 (2010).
\bibitem{MicrostripFlip}S. Michotte. Appl. Phys. Lett. {\bf 94}, 122512 (2009).
\bibitem{HistFitting}J. Lisenfeld, A. Lukashenko, and A. V. Ustinova. Appl. Phys. Lett. {\bf 91}, 232502 (2007).
\bibitem{Simmonds2004}R.W. Simmonds, K.M. Lang, D.A. Hite, S. Nam, D.P. Pappas, and John M. Martinis. Phys. Rev. Lett. {\bf 93}, 077003 (2004).
\bibitem{Cooper2004}K.B. Cooper, Matthias Stefen, R. McDermott, R.W. Simmonds, Seongshik Oh, D.A. Hite, D.P. Pappas, and John M. Martinis. Phys. Rev. Lett. {\bf 93} 180401 (2004).
\bibitem{Shane}M.S. Allman, F. Altomare, J.D. Whittaker, K. Cicak, D. Li, A. Sirois, J. Strong, J.D. Teufel, and R.W. Simmonds. Phys. Rev. Lett. {\bf 104} 177004 (2010).
\end{references}
\end{document}